\documentclass[12pt]{article}

\usepackage{graphicx}
\usepackage{amsfonts}
\usepackage{amsmath}

\def\gtwid{\mathrel{\raise.3ex\hbox{$>$\kern-.75em\lower1ex\hbox{$\sim$}}}}
\def\ltwid{\mathrel{\raise.3ex\hbox{$<$\kern-.75em\lower1ex\hbox{$\sim$}}}}
\def\square{\kern1pt\vbox{\hrule height 1.2pt\hbox{\vrule width 1.2pt\hskip 3pt
   \vbox{\vskip 6pt}\hskip 3pt\vrule width 0.6pt}\hrule height 0.6pt}\kern1pt}

\begin{document}

\begin{titlepage}

\begin{flushright}
UFIFT-QG-15-03
\end{flushright}

\vskip 1cm

\begin{center}
{\bf The Theorem of Ostrogradsky}
\end{center}

\vskip 1cm

\begin{center}
R. P. Woodard$^{\dagger}$
\end{center}

\vskip .5cm

\begin{center}
\it{Department of Physics, University of Florida,\\
Gainesville, FL 32611, UNITED STATES}
\end{center}

\vspace{.5cm}

\begin{center}
ABSTRACT
\end{center}

Ostrogradsky's construction of a Hamiltonian formalism for 
nondegenerate higher derivative Lagrangians is reviewed. The 
resulting instability imposes by far the most powerful restriction 
on fundamental, interacting, continuum Lagrangian field theories.
A discussion is given of the problems raised by attempts to evade
this restriction.

\begin{flushleft}
\end{flushleft}

\vskip 1cm

\begin{flushleft}
$^{\dagger}$ e-mail: woodard@phys.ufl.edu
\end{flushleft}

\end{titlepage}

\section{Introduction}

Albert Einstein famously commented, ``What really interests me is 
whether God had any choice in the creation of the world'' \cite{AE}.
Within the context of local Lagrangian field theory the answer seems 
to be that powerful restrictions exist but some freedom still remains 
regarding the choice of dynamical variables and symmetries. By far the
greatest restriction is the obstacle to including higher time
derivatives which is implied by Ostrogradsky's construction of a
canonical formalism for nondegenerate higher derivative Lagrangians
\cite{MO}.

Mikhail Vasilevich Ostrogradsky lived from 1801 to 1862. He was born 
to a poor family of Ukrainian ethnicity in Pashennaya, which is now in 
Ukraine but was at that time part of the vast Russian Empire. These were 
momentous years for Russia, bracketed by its rise to become the 
predominant military power during the Napoleonic Wars, and its 
humiliating collapse before Britain and France during the Crimean War.
Russian society was riven by the struggle between the forces of 
reaction and reform. Indeed, Ostrogradsky was denied his doctorate at 
the University of Kharkov because the mathematics professor who had
examined him was considered insufficiently religious. Later on,
Ostrogradsky was placed under police surveillance at the start of his 
career in the Imperial Russian capital of St. Petersburg \cite{CR}. 

Ostrogradsky studied and worked in Paris from 1822 through 1827. He
knew the leading French mathematicians of the time, including Cauchy,
who paid off his debts and secured him a teaching job. In 1826
Ostrogradsky stated and proved the divergence theorem, which was 
later re-discovered by Gauss in the 1830's. Ostrogradsky paid a much 
shorter visit to Paris in 1830. However, most of his professional life 
was spent in St. Petersburg where he was elected to the Imperial Academy 
of Sciences and played an important role in the teaching of mathematical
sciences. Ostrogradsky wrote in French and Russian \cite{CR}.

Ostrogradsky's higher derivative generalization of Hamilton's construction
was published in 1850 \cite{MO}. Ostrogradsky's construction implies that 
there is a linear instability in the Hamiltonians associated with Lagrangians 
which depend upon more than one time derivative in such a way that the 
higher derivatives cannot be eliminated by partial integration. This is 
probably why Newton was right to assume the laws of physics take the form of 
second differential equations when expressed in terms of fundamental dynamical 
variables. 

It might seem curious that Ostrogradsky did not appreciate the importance of 
his construction to fundamental theory. However, one must recall that the
researchers of his time were just beginning to make the connection between 
energy functionals and the concept of stability --- which in those days meant 
the absence of growing perturbations. The notion of quantum fluctuations 
exploring all perturbations was decades away, and the key insight that all 
dynamics is described by interacting continuum field theories was even 
further in the future.

Section presents Ostrogradsky's construction in the context of point 
particle whose position is $x(t)$. Section 3 discusses the consequences of 
this result for fundamental theory. Sections 4 and 5 deal with quantization
and degeneracy, respectively. Section 6 contains some concluding remarks.

\section{The Construction of Ostrogradsky}

This section presents Ostrogradsky's construction. First, the usual case of 
a first derivative Lagrangian is reviewed to fix concepts and notation. Then 
the case of second derivatives is presented. The section closes with a review 
of the general case of $N$ time derivatives.

\subsection{Hamilton's Construction}

In the usual case of $L = L(x,\dot{x})$, the Euler-Lagrange equation is,
\begin{equation}
\frac{\partial L}{\partial x} - \frac{d}{dt} \frac{\partial L}{\partial
\dot{x}} = 0 \; . \label{ELE1}
\end{equation}
The assumption that $\frac{\partial^2 L}{\partial \dot{x}^2} \neq 0$ is known 
as {\it nondegeneracy}. If the Lagrangian is nondegenerate one can write 
(\ref{ELE1}) in the form Newton assumed so long ago for the laws of physics,
\begin{equation}
\ddot{x} = \mathcal{F}(x,\dot{x}) \qquad \Longrightarrow \qquad x(t) =
\mathcal{X}(t,x_0,\dot{x}_0) \; . \label{newt}
\end{equation}
From this form it is apparent that solutions depend upon two pieces of initial
value data: $x_0 = x(0)$ and $\dot{x}_0 = \dot{x}(0)$. 

The fact that solutions require two pieces of initial value data means that
there must be two canonical coordinates, $X$ and $P$. They are traditionally
taken to be,
\begin{equation}
X \equiv x \qquad {\rm and} \qquad P \equiv \frac{\partial L}{\partial \dot{x}}
\; . \label{ctrans}
\end{equation}
The assumption of nondegeneracy implies one can invert the phase space
transformation (\ref{ctrans}) to solve for $\dot{x}$ in terms of $X$ and $P$. 
That is, there exists a velocity $V(X,P)$ such that,
\begin{equation}
\frac{\partial L}{\partial \dot{x}} \Biggl\vert_{x = X \atop \dot{x} = V}
= P \; . \label{invct}
\end{equation}

The canonical Hamiltonian is obtained by Legendre transforming on $\dot{x}$,
\begin{eqnarray}
H(X,P) & \equiv & P \dot{x} - L \; , \\
& = & P V(X,P) - L\Bigl(X,V(X,P)\Bigr) \; .
\end{eqnarray}
It is easy to check that the canonical evolution equations reproduce the 
inverse phase space transformation (\ref{invct}) and the Euler-Lagrange 
equation (\ref{ELE1}),
\begin{eqnarray}
\dot{X} & \equiv & \frac{\partial H}{\partial P} = V + P \frac{\partial V}{
\partial P} - \frac{\partial L}{\partial \dot{x}} \frac{\partial V}{\partial P}
= V \; , \\
\dot{P} & \equiv & -\frac{\partial H}{\partial X} = -P \frac{\partial V}{
\partial X} + \frac{\partial L}{\partial x} + \frac{\partial L}{\partial
\dot{x}} \frac{\partial V}{\partial X} = \frac{\partial L}{\partial x} \; .
\end{eqnarray}
This is the meaning of the statement, {\it the Hamiltonian generates time
evolution.} When the Lagrangian has no explicit time dependence, $H$ is also
the associated conserved quantity. Hence it possesses the key properties 
physicists want for the energy, and is unique up to canonical 
transformations.

A familiar example is the simple harmonic oscillator of mass $m$ and 
frequency $\omega$ whose Lagrangian is,
\begin{equation}
L = \frac12 m \dot{x}^2 - \frac12 m \omega^2 x^2 \; . \label{SHO}
\end{equation}
The equation of motion and its general solution are,
\begin{equation}
\ddot{x}(t) = -\omega^2 x(t) \qquad \Longrightarrow \qquad x(t) = x_0
\cos(\omega t) + \frac{ \dot{x}_0}{\omega} \sin(\omega t) \; . 
\label{freesol}
\end{equation}
The canonical variables for this system are,
\begin{equation}
X = x \qquad {\rm and} \qquad P = m \dot{x} \qquad \Longrightarrow \qquad
V(X,P) = \frac{P}{m} \; .
\end{equation}
And the Hamiltonian is,
\begin{equation}
H = \frac{P^2}{2 m} + \frac12 m \omega^2 X^2 \; .
\end{equation}
Because it is quadratic in both $X$ and $P$, the Hamiltonian $H(X,P)$ is
bounded below by zero.

\subsection{Ostrogradsky's Construction for Two Derivatives}

Now consider a system whose Lagrangian $L(x,\dot{x},\ddot{x})$ depends 
nonde\-gen\-er\-ate\-ly upon $\ddot{x}$. The Euler-Lagrange equation is,
\begin{equation}
\frac{\partial L}{\partial x} - \frac{d}{dt} \frac{\partial L}{\partial 
\dot{x}} + \frac{d^2}{dt^2} \frac{\partial L}{\partial \ddot{x}} = 0 \; .
\label{ELE2}
\end{equation}
Now nondegeneracy means $\frac{\partial^2 L}{\partial \ddot{x}^2} \neq 0$,
which implies that the Euler-Lagrange equation (\ref{ELE2}) can be cast in 
a form radically different from Newton's,
\begin{equation}
\ddddot{x} = \mathcal{F}(x,\dot{x},\ddot{x},\dddot{x}) \qquad 
\Longrightarrow \qquad x(t) = \mathcal{X}(t,x_0,\dot{x}_0,\ddot{x}_0,
\dddot{x}_0) \; .
\end{equation}

Because solutions now depend upon four pieces of initial value data there must
be four canonical coordinates. Ostrogradsky's choices for these are,
\begin{eqnarray}
X_1 \equiv x \qquad & , & \qquad P_1 \equiv \frac{\partial L}{\partial \dot{x}}
- \frac{d}{dt} \frac{\partial L}{\partial \ddot{x}} \; , \label{ct1} \\
X_2 \equiv \dot{x} \qquad & , & \qquad P_2 \equiv \frac{\partial L}{\partial 
\ddot{x}} \; . \label{ct2}
\end{eqnarray}
The assumption of nondegeneracy implies one can invert the phase space
transformation (\ref{ct1}-\ref{ct2}) to solve for $\ddot{x}$ in terms of
$X_1$, $X_2$ and $P_2$. That is, there exists an acceleration $A(X_1,X_2,P_2)$
such that,
\begin{equation}
\frac{\partial L}{\partial \ddot{x}} \Biggl\vert_{{x = X_1 \atop \dot{x} =
X_2} \atop \ddot{x} = A} = P_2 \; . \label{invct2}
\end{equation}
Note that the acceleration $A(X_1,X_2,P_2)$ does nor depend upon $P_1$. The
momentum $P_1$ is only needed for the third time derivative.

Ostrogradsky's Hamiltonian is obtained by Legendre transforming on
$\dot{x} = x^{(1)}$ and $\ddot{x} = x^{(2)}$,
\begin{eqnarray}
\lefteqn{H(X_1,X_2,P_1,P_2) \equiv \sum_{i=1}^2 P_i x^{(i)} - L \; , } \\
& & = P_1 X_2 + P_2 A(X_1,X_2,P_2) - L\Bigl(X_1,X_2,A(X_1,X_2,P_2)\Bigr) \; .
\label{Host}
\end{eqnarray}
The time evolution equations are those suggested by the notation,
\begin{equation}
\dot{X_i} \equiv \frac{\partial H}{\partial P_i} \qquad {\rm and} \qquad 
\dot{P}_i \equiv - \frac{\partial H}{\partial X_i} \; .
\end{equation}
To check that they generate time evolution, note first that the evolution 
equation for $X_1$ is,
\begin{equation}
\dot{X}_1 = \frac{\partial H}{\partial P_1} = X_2 \; .
\end{equation}
Of course this reproduces the phase space transformation $\dot{x} = X_2$ in 
(\ref{ct2}). The evolution equation for $X_2$ similarly reproduces 
(\ref{invct2}),
\begin{equation}
\dot{X}_2 = \frac{\partial H}{\partial P_2} = A + P_2 \frac{\partial A}{
\partial P_2} - \frac{\partial L}{\partial \ddot{x}} \frac{\partial A}{\partial
P_2} = A \; .
\end{equation}
The phase space transformation $P_1 = \frac{\partial L}{\partial
\dot{x}} - \frac{d}{dt} \frac{\partial L}{\partial \ddot{x}}$ (\ref{ct1})
comes from the evolution equation for $P_2$,
\begin{equation}
\dot{P}_2 = -\frac{\partial H}{\partial X_2} = -P_1 - P_2 \frac{\partial A}{
\partial X_2} + \frac{\partial L}{\partial \dot{x}} + \frac{\partial L}{
\partial \ddot{x}} \frac{\partial A}{\partial X_2} = -P_1 + \frac{\partial L}{
\partial \dot{x}} \; .
\end{equation}
And the Euler-Lagrange equation (\ref{ELE2}) follows from the evolution 
equation for $P_1$,
\begin{equation}
\dot{P}_1 = -\frac{\partial H}{\partial X_1} = -P_2 \frac{\partial A}{\partial 
X_1} + \frac{\partial L}{\partial x} + \frac{\partial L}{\partial \ddot{x}}
\frac{\partial A}{\partial X_1} = \frac{\partial L}{\partial x} \; .
\end{equation}
Hence Ostrogradsky's Hamiltonian generates time evolution. When the Lagrangian 
contains no explicit dependence upon time it is also the conserved Noether 
current. It is therefore the energy, again up to canonical transformation.

Ostrogradsky's Hamiltonian (\ref{Host}) is linear in the canonical momentum 
$P_1$, which means that no system of this form can be stable. In fact, there 
is not even any barrier to decay. Note the power and generality of the result: 
it applies to every Lagrangian $L(x,\dot{x},\ddot{x})$ which depends 
nondegenerately upon $\ddot{x}$, independent of the details. The only 
assumption is nondegeneracy, and that simply means one cannot eliminate 
$\ddot{x}$ by partial integration.

It is useful to consider a higher derivative example which depends upon a 
dimensionless parameter $\epsilon$ that quantifies its deviation from the 
simple harmonic oscillator (\ref{SHO}),
\begin{equation}
L = -\frac{\epsilon m}{2 \omega^2} \ddot{x}^2 + \frac{m}2 \dot{x}^2
- \frac{m\omega^2}2 x^2 \; . \label{HDO}
\end{equation}
The Euler-Lagrange equation and its general solution are,
\begin{eqnarray}
0 & = & -m \Bigl[ \frac{\epsilon}{\omega^2} \ddddot{x} + \ddot{x} + 
\omega^2 x\Bigr] \; , \label{HDE} \\
x(t) & = & C_+ \cos(k_+ t) + S_+ \sin(k_+ t) + C_- \cos(k_- t) + S_- 
\sin(k_- t) \; . \label{gensol} \qquad
\end{eqnarray}
Here the two frequencies are,
\begin{equation}
k_{\pm} \equiv \omega \sqrt{ \frac{1 \mp \sqrt{1 \!-\! 4 \epsilon}}{2 
\epsilon} } \; , \label{kpm}
\end{equation}
and the constants $C_{\pm}$ and $S_{\pm}$ are functions of the initial value data,
\begin{eqnarray}
C_+ = \frac{k_-^2 x_0 \!+\! \ddot{x}_0}{k_-^2 \!-\! k_+^2} \qquad & , & \qquad
S_+ = \frac{k_-^2 \dot{x}_0 \!+\! \dddot{x}_0}{k_+ (k_-^2 \!-\! k_+^2)} \; , \\
C_- = \frac{k_+^2 x_0 \!+\! \ddot{x}_0}{k_+^2 \!-\! k_-^2} \qquad & , & \qquad
S_- = \frac{k_+^2 \dot{x}_0 \!+\! \dddot{x}_0}{k_- (k_+^2 \!-\! k_-^2)} \; .
\end{eqnarray}
For this model Ostrogradsky's two conjugate momenta (\ref{ct1}-\ref{ct2}) are,
\begin{eqnarray}
P_1 = m \dot{x} + \frac{\epsilon m}{\omega^2} \dddot{x} \qquad & \Leftrightarrow & 
\qquad \dddot{x} = \frac{\omega^2 P_1 \!-\! m \omega^2 X_2}{\epsilon m} \; , \\
P_2 = - \frac{\epsilon m}{\omega^2} \ddot{x} \qquad & \Leftrightarrow & 
\qquad \ddot{x} \equiv A = -\frac{\omega^2 P_2}{\epsilon m} \; .
\end{eqnarray}
The Hamiltonian can be expressed alternatively in terms of canonical variables,
configuration space variables, or the constants $C_{\pm}$ and $S_{\pm}$,
\begin{eqnarray}
H & = & P_1 X_2 - \frac{\omega^2}{2 \epsilon m} P_2^2 - \frac{m}2 X_2^2 + 
\frac{m\omega^2}2 X_1^2 \; , \label{H1} \\
& = & \frac{\epsilon m}{\omega^2} \dot{x} \dddot{x} - \frac{\epsilon m}{2 \omega^2}
\ddot{x}^2 + \frac{m}2 \dot{x}^2 + \frac{m \omega^2}2 x^2 \; , \label{H2} \\
& = & \frac{m}2 \sqrt{1 \!-\! 4 \epsilon} \, k_+^2 (C_+^2 \!+\! S_+^2) - 
\frac{m}2 \sqrt{1 \!-\! 4 \epsilon} \, k_-^2 (C_-^2 \!+\! S_-^2) \; . \label{H3}
\end{eqnarray}
The last form (\ref{H3}) makes it clear that the $+$ modes carry positive 
energy and the $-$ modes carry negative energy. 

\subsection{Ostrogradsky's Construction for $N$ Derivatives}

Consider a Lagrangian $L\left(x,\dot{x},\dots,x^{(N)}\right)$ which depends 
upon the first $N$ derivatives of $x(t)$. If this Lagrangian depends 
nondegenerately upon the $N$-th derivative $x^{(N)}$ then the Euler-Lagrange 
equation is linear in the $2N$-th derivative $x^{(2N)}$,
\begin{equation}
\sum_{i=0}^N \left(-{d \over dt}\right)^i {\partial L \over \partial x^{(i)}}
= 0 \; . \label{ELEN}
\end{equation}
The canonical phase space must therefore possess $2N$ coordinates which 
Ostrogradsky chooses to be,
\begin{equation}
X_i \equiv x^{(i-1)} \qquad {\rm and} \qquad P_i \equiv \sum_{j=i}^N \Bigl(-
\frac{d}{dt}\Bigr)^{j-i} \frac{\partial L}{\partial x^{(j)}} \; . \label{Ncts}
\end{equation}
Nondegeneracy means one can solve for $x^{(N)}$ in terms of $P_N$ and the
$X_i$'s. That is, there exists a function $\mathcal{A}(X_1,\ldots,X_N,P_N)$
such that,
\begin{equation}
\frac{\partial L}{\partial q^{(N)}} \Biggl\vert_{x^{(i-1)} = X_i \atop
x^{(N)} = \mathcal{A}} = P_N \; . \label{nondeg}
\end{equation}

For general $N$ Ostrogradsky's Hamiltonian takes the form,
\begin{eqnarray}
H & \equiv & \sum_{i=1}^N P_i x^{(i)} - L \; , \\
& = & P_1 X_2 + P_2 X_3 + \cdots + P_{N-1} X_N + P_N \mathcal{A} -
L\Bigl(X_1,\ldots,X_N,\mathcal{A}\Bigr) \; . \label{HN}
\end{eqnarray}
The evolution equations are,
\begin{equation}
\dot{X}_i \equiv \frac{\partial H}{\partial P_i} \qquad {\rm and} \qquad 
\dot{P}_i \equiv -\frac{\partial H}{\partial X_i} \; .
\end{equation}
It is simple to check that these evolution equations reproduce the 
canonical relations (\ref{Ncts}) and the Euler-Lagrange equation 
(\ref{ELEN}). The first $(N-1)$ equations for $X_i$ verify the
definition of $X_{i+1}$,
\begin{equation}
i=1,\dots,(N-1) \qquad \Longrightarrow \qquad \dot{X}_i = X_{i+1} \; .
\end{equation}
The evolution equation for $X_N$ is similar,
\begin{equation}
\dot{X}_N = \mathcal{A} + P_N \frac{\partial \mathcal{A}}{\partial P_N}
- \frac{\partial L}{\partial x^{(N)}} \frac{\partial \mathcal{A}}{
\partial P_N} = \mathcal{A} \; .
\end{equation}
The last $(N-1)$ equations for $P_i$ reproduce the definition of 
$P_{i-1}$,
\begin{eqnarray}
i=2,\dots,N \quad \Longrightarrow \quad \dot{P}_i & = & -P_{i-1} - P_N
\frac{\partial \mathcal{A}}{\partial X_i} + \frac{\partial L}{\partial
x^{(i-1)}} + \frac{\partial L}{\partial x^{(N)}} \frac{\partial 
\mathcal{A}}{\partial X_i} \; , \qquad \\
& = & -P_{i-1} + \frac{\partial L}{\partial x^{(i-1)}} \; .
\end{eqnarray}
And the evolution equation for $P_1$ gives the Euler-Lagrange eqaution
(\ref{ELEN}),
\begin{equation}
\dot{P}_1 = -P_N \frac{\partial \mathcal{A}}{\partial X_1} + 
\frac{\partial L}{\partial x} + \frac{\partial L}{\partial x^{(N)}}
\frac{\partial \mathcal{A}}{\partial X_1} = 
\frac{\partial L}{\partial x} \; .
\end{equation}
Hence (\ref{HN}) generates time evolution. It is also the Noether current 
for the case where the Lagrangian contains no explicit time dependence.
The Hamiltonian (\ref{HN}) is therefore what any physicist would call the 
energy, up to canonical transformation.

The Hamiltonian (\ref{HN}) is linear in $P_1, P_2, \ldots P_{N-1}$. Only 
with respect to $P_N$ might it be bounded from below. For large $N$ the 
fraction of linear directions approaches $\frac12$, so adding more higher 
derivatives makes the instability worse rather than better.

\section{Nature of the Instability}

Ostrogradsky's result implies that the Hamiltonian of a nondegenerate 
higher derivative theory is unbounded below, and also above. This
section discusses the manner in which the instability manifests, and
what it implies for fundamental theory. Six short subsections make
the points:
\begin{enumerate}
\item{The Ostrogradskian instability drives the dynamical variable to
a special kind of time dependence, not a special numerical value.}
\item{The same Ostrogradskian dynamical variable carries both positive 
and negative energy creation and annihilation operators.}
\item{If a system which suffers from the Ostrogradskian instability
interacts, then the empty state can decay into a collection of positive 
and negative energy excitations.}
\item{If a system which suffers from the Ostrogradskian instability is 
a continuum field theory, the vast entropy at infinite 3-momentum will 
make the decay instantaneous.}
\item{For interacting systems which suffer from the Ostrogradskian
instability, degrees of freedom with large 3-momentum do not decouple 
from low energy physics.}
\item{The imposition of a single, global constraint on the energy 
functional does not ameliorate the Ostrogradskian instability.}
\end{enumerate}

\subsection{Kinetic Instability}

Physicists are familiar with instabilities of the potential energy. 
In this case energy is released as the dynamical variable approaches 
some special value. The Ostrogradskian instability is instead a problem 
with the kinetic energy, and it manifests by the dynamical variable 
developing a special time dependence. Checking that the energy is
bounded below for constant values of the dynamical variable in no way
establishes that a system is free of the Ostrogradskian instability.
Consider, for example, the higher derivative oscillator (\ref{HDO}).
Expression (\ref{H2}) shows that its energy is bounded below by zero
for any constant value of $x(t)$. Negative energies are attained by 
making $\ddot{x}(t)$ large and/or making $\dddot{x}(t)$ large while 
keeping the combination $\dot{x}(t) \!+\! \frac{\epsilon}{\omega^2}
\dddot{x}(t)$ fixed.

\subsection{Double Duty for Dynamical Variables}

Physicists are used to resolving linearized dynamical variables into
creation and annihilation operators. For the harmonic oscillator
solution (\ref{freesol}) this is done by using the Euler relation to
identify a lowering operator proportional to $e^{-i\omega t}$ and a
raising operator proportional to $e^{i \omega t}$,
\begin{equation}
 x_0 \cos(\omega t) + \frac{ \dot{x}_0}{\omega} \sin(\omega t) =
\frac12 \Bigl[ x_0 + \frac{i}{\omega} \dot{x}_0\Bigr] e^{-i \omega t}
+ \frac12 \Bigl[ x_0 - \frac{i}{\omega} \dot{x}_0\Bigr] e^{i \omega t}
\; .
\end{equation}
The usual rule is that each dynamical variable harbors either zero or one
set of creation and annihilation operators at linearized order. From 
expression (\ref{gensol}) one can see that the same higher derivative
dynamical variable carries both positive and negative energy creation and
annihilation operators. This means that local interactions which involve 
the dynamical variable necessarily couple the two sectors.

\subsection{The Vacuum Can Decay}

Now consider an interacting, continuum field theory which possesses the
Ostrogradskian instability. In particular consider its likely particle
spectrum about some ``empty'' solution in which the field is constant.
Because the Hamiltonian is linear in all but one of the conjugate momenta
it is possible to arbitrarily increase or decrease the energy by moving 
different directions in phase space. Hence there must be both positive 
energy and negative energy particles --- just as there are in the higher 
derivative oscillator (\ref{HDO}). As in that point particle model, the 
same continuum field must carry the creation and annihilation operators 
of both the positive and the negative energy particles. If the theory is 
interacting at all --- that is, if its Lagrangian contains a higher than 
quadratic power of the field --- then there will be interactions between 
positive and negative energy particles. Depending upon the interaction, 
the empty state can decay into some collection of positive and negative 
energy particles. 

\subsection{Entropy Drives Vacuum Decay}

Recall the reason that excited states of atoms decay in nature. It is 
certainly not to reduce the energy of the full system --- including the
interaction with electromagnetism --- but rather to redistribute the
constant total energy into the largest possible class of states. There 
is one way for the atom not to decay, compared with an infinite number of 
ways the atom can decay and emit one or more recoil photons. Note also
that explicit computations of the decay time employ vacuum fluctuations
of the electromagnetic field to provide the necessary perturbation.

Atomic decays have just the fixed energy difference between the two 
states to apportion, so they are chiefly driven by the arbitrary
directions which can be taken by the decay products. In contrast, the
decay of an interacting, nondegenerate higher derivative field theory
can involve particles of {\it any} energy, as long as the total
sums to zero. So one should think of the decay rate as having the same 
sort of angular factors as an atomic decay at some fixed energy, 
followed by one or more integrals --- all the way to infinity --- over 
the magnitudes of the various energies. The volume of phase space is so 
large that these integrations cause the decay to be instantaneous. 
Indeed, the only way people derive finite decay rates for particles 
with a kinetic instability is by cutting off the phase space at some 
point, in which case the rate is dominated by the cutoff, for example 
\cite{CJM}. Such a cutoff might make sense if the kinetic instability 
appeared in some nonlocal effective field theory, but it has no place
in fundamental physics.

Note that the decay does not just happen once. It is even more 
entropicly favored for there to be two decays, and better yet 
for more. In fact the system instantly evaporates into a maelstrom 
of positive and negative energy particles. Whether or not such a state
has a proper mathematical representation, it certainly does not describe
the universe of human experience in which all particles have positive 
energy and empty space remains empty.

Note also that this conclusion only follows if the higher derivative 
theory possesses both interactions and continuum particles. The point 
particle oscillator (\ref{HDO}) has no interactions, so its negative 
energy degree of freedom is unobservable. However, it is conceivable  
that this higher derivative oscillator could be coupled to a discrete 
system without engendering any instability. The feature which drives 
explosive vacuum decay is the vast entropy of phase space. Without 
that it becomes an open question whether or not there is anything wrong 
with a higher derivative theory. Of course the physical universe seems 
to be described by continuum field theory down to at least $2.8 \times 
10^{-19}$ meters \cite{DB}, and any observable degree of freedom must 
interact, or else it could not be observed, so these seem to be safe 
assumptions.

\subsection{Large $\Vert \vec{k}\Vert$ Modes Do Not Decouple}

Physicists are used to ignoring very high energy modes, except for 
renormalizations of low energy parameters. This procedure is quite correct
for positive energy modes in a stable theory because exciting a mode
requires energy which must be drawn from de-exciting other modes, and 
any given state only has some fixed amount of energy. However, that
justification fails for a theory which suffers from the Ostrogradskian 
instability because even a very high (positive or negative) energy
mode can be excited by also exciting modes with the opposite energy.
Instead of these large $k$ modes decoupling, they couple ever more 
strongly as $k$ grows, because more and more ways open up to balance
its energy by exciting lower modes of the opposite sign.

\subsection{Constraints on $H$ Accomplish Nothing}

It is sometimes imagined that the energy of a higher derivative theory decays
with time. That is not true. Provided one is dealing with a complete system,
and provided there is no external time dependence, the energy of a higher
derivative system is conserved, just as it would be under those conditions
for a lower derivative theory. This conservation is apparent for the higher 
derivative oscillator (\ref{HDO}) from expression (\ref{H3}). 

The physical problem with nondegenerate higher derivative theories is not that
their energies decay to lower and lower values. The problem is rather that 
certain sectors of the theory become arbitrarily highly excited when one is 
dealing with an interacting, continuum field theory which has nondegenerate 
higher derivatives. For example, Boulware, Horowitz and Strominger \cite{BHS} 
showed that the energy is zero for any asymptotically flat solution of the 
higher derivative field equations derived from the Lagrangian,
\begin{equation}
\mathcal{L} = -\frac14 \alpha C_{\mu\nu\rho\sigma} C^{\mu\nu\rho\sigma} 
\sqrt{-g} -\frac14 \beta R^2 \sqrt{-g} \; , \label{zeroE}
\end{equation}
where $C_{\mu\nu\rho\sigma}$ is the Weyl tensor and $R$ is the Ricci scalar.
However, this model is still unstable for $\alpha \neq 0$, as its creators
realized.\footnote{It is also worth noting that the requirement of asymptotic 
flatness in this model would preclude the response to normal matter, and that
imposing the correct asymptotic condition gives rise to nonzero energy
\cite{DT}.} 

\section{Quantization}

Quantization is very important to understanding the Ostrogradskian instability
because 0-point fluctuations provide the perturbations needed to ensure that
the potential for vacuum decay is actually realized. However, a quantum higher
derivative system has some peculiarities. For example, it is obvious from 
relations (\ref{ct1}-\ref{ct2}) that position and velocity commute! Further,
the wave function of a higher derivative theory depends upon position and 
velocity. This section argues first that the classical instability survives
{\it canonical} quantization. After presenting a worked-out example, the 
curious noncanonical quantizations which sometimes appear in the literature
are discussed.

\subsection{A Large Phase Space Instability}

It is often imagined that quantization might protect a higher derivative 
system against the Ostrogradskian instability the same way that quantization
prevents the collapse of atoms coupled to electromagnetism. This is a failure 
to understand correspondence limits. In the Heisenberg picture the equations
of classical mechanics are identical to those of quantum mechanics. It also
means the very same thing to solve these equations: one expresses the 
dynamical variable in terms of time and the allowed initial value data, as
in expressions (\ref{freesol}) and (\ref{gensol}). The only difference 
between classical and quantum mechanics is that the classical initial value
data are numbers which can take any value whereas the quantum initial value 
data include noncommuting conjugate operators which obey the Uncertainty 
Principle. The only classical phenomena that can be affected by quantization
are those whose realization requires localizing conjugate variables to some 
volume of the classical phase space smaller than $\hbar$. So quantum atoms
are stable because localizing the electrons too near the nucleus necessarily
induces a large kinetic energy.

In contrast, the Ostrogradskian instability derives from the fact that 
$P_1 X_2$ can be made arbitrarily negative by taking $P_1$ either very negative,
for positive $X_2$, or else very positive, for negative $X_2$. {\it This covers 
essentially half the classical phase space!} Further, the variables $X_2$ and 
$P_1$ commute with one another in Ostrogradskian quantum mechanics. So there
is no reason to expect that the Ostrogradskian instability is unaffected by 
quantization.

\subsection{Quantum Higher Derivative Oscillator}

Consider the second derivative oscillator (\ref{HDO}) discussed in section 2.2.
There can be no ground state in the presence of the Ostrogradskian instability 
but one might define an ``empty'' state wavefunction, $\Omega(X_1,X_2)$ which 
has the minimum excitation in both the positive and negative energy degrees of 
freedom. The procedure for doing this is simple: first identify the positive
and negative energy lowering operators $\alpha_{\pm}$, and then solve the
equations,
\begin{equation}
\alpha_+ \vert \Omega \rangle = 0 = \alpha_- \vert \Omega \rangle \; .
\label{wavef}
\end{equation}
One can recognize the raising and lowering operators by expressing the 
general solution (\ref{gensol}) in terms of exponentials,
\begin{eqnarray}
\lefteqn{x(t) = \frac12 (C_+ \!+\! i S_+) e^{-ik_+ t} + \frac12 (C_+ \!-\! 
i S_+) e^{ik_+ t} } \nonumber \\
& & \hspace{2cm} + \frac12 (C_- \!+\! i S_-) e^{-ik_-t } + \frac12 (C_- \!-\! 
i S_-) e^{ik_- t} \; .
\end{eqnarray}
Recall that the $k_+$ mode carries positive energy, so its lowering operator
must be proportional to the $e^{-ik_+ t}$ term,
\begin{eqnarray}
\alpha_+ & \sim & C_+ + i S_+ \; , \\
& \sim & \frac{m k_+}2 \Bigl(1 \!+\! \sqrt{1 \!-\! 4\epsilon}\Bigr) X_1 + 
i P_1 - k_+ P_2 - \frac{i m}2 \Bigl(1 \!-\! \sqrt{1 \!-\! 4\epsilon}\Bigr) 
X_2 \; .
\end{eqnarray}
The $k_-$ mode carries negative energy, so its lowering operator must be 
proportional to the $e^{+i k_- t}$ term,
\begin{eqnarray}
\alpha_- & \sim & C_- - i S_- \; , \\
& \sim & \frac{m k_-}2 \Bigl(1 \!-\! \sqrt{1 \!-\! 4\epsilon}\Bigr) X_1 - 
i P_1 - k_- P_2 + \frac{i m}2 \Bigl(1 \!+\! \sqrt{1 \!-\! 4\epsilon}\Bigr) 
X_2 \; .
\end{eqnarray}
Writing $P_i = -i\hbar \frac{\partial}{\partial X_i}$ reveals that the unique 
solution to (\ref{wavef}) has the form,
\begin{equation}
\Omega(X_1,X_2) = N \exp\Biggl[-\frac{m \sqrt{1 \!-\! 4\epsilon}}{2 \hbar(k_+ 
\!+\! k_-)} \Bigl(k_+ k_- X_1^2 + X_2^2\Bigr) - \frac{i \sqrt{\epsilon} m}{\hbar} 
X_1 X_2\Biggr] \; . \label{true}
\end{equation}

The empty wave function (\ref{true}) is obviously normalizable, so it gives a
state of the quantum system. One can build a complete set of normalized 
stationary states by acting arbitrary numbers of $+$ and $-$ raising operators 
on it,
\begin{equation}
\vert N_+ , N_-\rangle \equiv \frac{(\alpha_+^{\dagger})^N_+}{\sqrt{N_+ !}}
\frac{(\alpha_-^{\dagger})^N_-}{\sqrt{N_- !}} \vert \Omega \rangle \; .
\end{equation}
On this space of states the Hamiltonian operator is unbounded below, just as
in the classical theory,
\begin{equation}
H \vert N_+ , N_- \rangle = \hbar \Bigl(N_+ k_+ - N_- k_-\Bigr) \vert N_+ , N_-
\rangle \; .
\end{equation}
This is the correct way to quantize a higher derivative theory. One evidence
of this fact is that classical configurations of negative energy correspond to 
quantum negative energy states.

\subsection{Unitarity versus Instability}

Particle physicists who quantize higher derivative theories do not typically
recognize a problem with stability; they instead discuss a breakdown of
unitarity, for example \cite{KS}. This is accomplished by regarding the 
negative energy lowering operator as a positive energy raising operator. So 
one defines a ``ground state'' $\vert \overline{\Omega} \rangle$ which obeys 
the equations,
\begin{equation}
\alpha_+ \vert \overline{\Omega} \rangle = 0 = \alpha_-^{\dagger} \vert 
\overline{\Omega} \rangle \; . \label{falsewave}
\end{equation}
The unique wave function which solves these equations is,
\begin{equation}
\overline{\Omega}(X_1,X_2) = N \exp\Biggl[-\frac{m \sqrt{1 \!-\! 
4\epsilon}}{2 \hbar (k_- \!-\! k_+)} \Bigl(k_+ k_- X_1^2 - X_2^2\Bigr) + 
\frac{i \sqrt{\epsilon} m}{\hbar} X_1 X_2 \Biggr] \; . \label{wrong}
\end{equation}
The wave function (\ref{wrong}) is not normalizable, so it does 
not correspond to a state of the quantum system \cite{HH}. However, 
particle physicists define a formal ``space of states'' based upon 
$\vert\overline{\Omega} \rangle$,
\begin{equation}
\vert \overline{N_+ , N_-}\rangle \equiv \frac{(\alpha_+^{\dagger})^{N_+}}{
\sqrt{N_+ !}} \frac{(\alpha_-)^{N_-}}{\sqrt{N_- !}} \vert \overline{\Omega} 
\rangle \; .
\end{equation}
Although these wave functions are no more normalizable than 
$\overline{\Omega}(X_1,X_2)$, they are all positive energy eigenfunctions,
\begin{equation}
H \vert \overline{N_+ , N_-} \rangle = \hbar \Bigl(N_+ k_+ + N_- k_-\Bigr) 
\vert \overline{N_+ , N_-} \rangle \; .
\end{equation}

The problem with unitarity emerges because $\vert \overline{\Omega}\rangle$ 
is defined to have unit norm, but the commutation relations are unchanged,
\begin{equation}
[\alpha_+,\alpha_+^{\dagger}] = 1 = [\alpha_-,\alpha_-^{\dagger}] \; .
\end{equation}
Hence the norm of any state with odd $N_-$ is negative. The first of these 
negative norm states is,
\begin{equation}
\langle \overline{0,1} \vert \overline{0,1}\rangle =
\langle \overline{\Omega} \vert \alpha_-^{\dagger} \alpha_- \vert \overline{
\Omega} \rangle = - \langle \overline{\Omega} \vert \overline{\Omega} \rangle
\; .
\end{equation}
The next step is to invoke the probabilistic interpretation of quantum 
mechanics which requires norms to be positive because probabilities are. 
Therefore, the negative norm states must be excised from the space of states. 
However, doing that results in a nonunitary S-matrix because scattering
processes inevitably mix positive and negative norm states, just as the 
correctly-quantized, indefinite-energy theory allows processes which mix 
positive and negative energy particles.

It is important to note that the potential for invoking noncanonical
quantization schemes to change the range of allowed energies is present 
even in the usual, first derivative systems. The Schrodinger equation 
$H \psi(X) = E \psi(X)$ is a second order differential equation, which 
possesses two linearly independent solutions for every value of the energy 
$E$. It is only by insisting upon normalizable wave functions that 
quantized energies emerge. Many other peculiar things happen if one 
abandons normalizability \cite{RPW1,TW0}. In particular, the 
Correspondence Principle fails, so that taking $\hbar$ to zero gives a
different classical system from the one which originally motivated the
analysis. That is the case for $PT$-symmetric quantizations of higher 
derivative systems \cite{BM,AM}.

\section{Degeneracy}

The only way anyone has ever found to avoid the Ostrogradskian 
instability is by violating the assumption of nondegeneracy upon 
which it is based. This section discusses three ways this can 
happen: through partial integration, through gauge invariance, 
and by imposing constraints by fiat \cite{CFLT}.

\subsection{Trivial Degeneracy}

The simplest form of degeneracy derives from adding a total derivative to a
first order system. Examples include the Hilbert action of general relativity, 
Lovelock gravity \cite{Love} and Galileons \cite{NRT,RT}. In that case one 
simply performs a partial integration, and discards the surface term to obtain 
a Lagrangian which contains only first time derivatives. For example, the 3rd 
Lagrangian for a scalar Galileon $\pi(t,\vec{x})$ reduces to first order form 
as,
\begin{eqnarray}
\partial_{\mu} \pi \partial^{\mu} \pi \partial^2 \pi & = & 
\frac{\partial}{\partial t} \Bigl[\frac13 \dot{\pi}^3 \!-\! \dot{\pi} 
\vec{\nabla} \pi \!\cdot\! \vec{\nabla} \pi\Bigr] + 2 \dot{\pi} 
\vec{\nabla} \pi \!\cdot\! \vec{\nabla} \dot{\pi} + \nabla^2 \pi 
\vec{\nabla} \pi \!\cdot\! \vec{\nabla} \pi \; , \qquad \\
& \longrightarrow & 2 \dot{\pi} \vec{\nabla} \pi \!\cdot\! \vec{\nabla} 
\dot{\pi} + \nabla^2 \pi \vec{\nabla} \pi \!\cdot\! \vec{\nabla} \pi \; .
\end{eqnarray}
Note that it is only necessary to eliminate higher {\it time} derivatives; 
there is no problem if the Lagrangian contains higher spatial derivatives,
or mixed first time and space derivatives.

\subsection{Gauge Degeneracy}

All theories which possess continuous symmetries are degenerate, irrespective 
of whether or not they possess higher derivatives. A familiar example is the 
relativistic point particle, whose dynamical variable is $X^{\mu}(\tau)$ and 
whose Lagrangian is,
\begin{equation}
L = -m \sqrt{-\eta_{\mu\nu} \dot{X}^{\mu} \dot{X}^{\nu}} \; .
\end{equation}
The conjugate momentum is,
\begin{equation}
P_{\mu} \equiv \frac{m \dot{X}_{\mu}}{\sqrt{-\dot{X}^2}} \; . \label{RPE}
\end{equation}
One cannot solve (\ref{RPE}) for $\dot{X}^{\mu}$ in terms of $X^{\mu}$ and
$P_{\mu}$ because the equation is homogeneous of degree zero. The continuous 
symmetry associated with this degeneracy is invariance under changes of the
parameter $\tau \longrightarrow \tau'$,
\begin{equation}
X^{\mu}(\tau) \longrightarrow X^{\prime \mu}(\tau) \equiv X^{\mu}\Bigl({\tau'
}^{-1}(\tau)\Bigr) \; .
\end{equation}

The cure for symmetry-induced degeneracy is simply to fix the symmetry by
imposing gauge conditions. Then the gauge-fixed Lagrangian should no longer 
be degenerate in terms of the remaining variables. For example, one might fix
the parameter $\tau$ to obey $\tau = X^0(\tau)$. In that case the gauge-fixed
particle Lagrangian is,
\begin{equation}
L_{\rm GF} = -m \sqrt{1 - \dot{\vec{X}} \cdot \dot{\vec{X}} } \; ,
\end{equation}
and the relations for the momenta are simple to invert,
\begin{equation}
P_i \equiv \frac{m \dot{X}_i}{\sqrt{1 - \dot{\vec{X}} \cdot \dot{\vec{X}} }} 
\qquad \Longleftrightarrow \qquad \dot{X}^i = \frac{P^i}{\sqrt{m^2 + \vec{P}
\cdot \vec{P}}} \; .
\end{equation}

When a continuous symmetry is used to eliminate a dynamical variable, the
equation of motion of this variable typically becomes a {\it constraint}. 
For symmetries enforced by means of a compensating field --- such as making
the Hilbert action local Lorentz invariant using the antisymmetric components 
of the vierbein \cite{RPW2}, or Weyl invariant using a scalar \cite{TW1} --- 
the associated constraints are tautologies of the form $0 = 0$. Sometimes the 
constraints are nontrivial, but implied by the equations of motion. An example 
of this kind is the relativistic particle considered above. In synchronous 
gauge ($\tau = X^0(\tau)$) the equation of the gauge-fixed zero-component 
implies that the Hamiltonian is conserved,
\begin{equation}
\frac{d}{d\tau} \Biggl( \frac{m \dot{X}_0}{\sqrt{-\eta_{\mu\nu} \dot{X}^{\mu} 
\dot{X}^{\nu}}} \Biggr) = 0 \longrightarrow \frac{d}{dt} \Bigl(
\sqrt{m^2 + \vec{P} \!\cdot\! \vec{P} } \Bigr) = 0 \; .
\end{equation}
And sometimes the constraints give nontrivial relations between the canonical 
variables that generate residual, time-independent symmetries. In this case
another degree of freedom can be removed. An example of this kind of constraint 
is Gauss' Law in temporal gauge electrodynamics. 

When constraints of the third type are present one must check whether or not 
they affect the instability. This obviously depends on the particular model
being studied but a necessary condition for avoiding the Ostrogradskian
instability is that the number of gauge constraints must equal or exceed the
number of unstable directions in the canonical phase space. Because the number 
of constraints for any given symmetry is fixed, whereas the number of unstable 
directions increases with the number of higher derivatives, it follows that
gauge constraints can at best avoid instability for some fixed number of higher
derivatives.

A good example of gauge degeneracy is provided by the quadratic curvature model
(\ref{zeroE}) which was exhibited at the end of section 3 to show the 
irrelevance of a global constraint on the Hamiltonian. As long as $\alpha$ and
$\beta$ are both nonzero, there are six independent, higher derivative momenta 
at each space point, whereas there are only four local constraints. If $\beta  
= 0$ the model acquires a new local symmetry --- Weyl invariance ---- which 
adds another local constraint. Hence there are either one or two unconstrained 
instabilities per space point for $\alpha \neq 0$. There are an infinite number 
of space points, so the addition of a single, global constraint does not change 
anything. 

The case of $\alpha = 0$ is special. If $\beta$ has the right sign the resulting
model has long been known to have positive energy \cite{AAS0,AS}. This result in 
no way contradicts the previous analysis. When $\alpha = 0$, the terms which 
carry second derivatives are contracted in such way that only a single component 
of the metric carries higher derivatives. So the counting is one unstable 
direction per space point versus four local constraints, which means the 
constraints can prevent the Ostrogradskian instability. 

\subsection{Imposed Degeneracy}

Many attempts to evade the Ostrogradskian instability are based on 
segregating higher derivatives to interaction terms so that the free theory 
possesses no extra solutions. This renders the instability invisible 
to perturbative scrutiny but does not avoid it. One can see from the 
construction of section 2 that the sole assumption needed to derive the 
instability is nondegeneracy, irrespective of how one organizes any 
approximation technique. On the other hand, there is a way of imposing 
constraints so as to make the theory agree with its perturbative 
development. When this is done there are no more higher derivative 
degrees of freedom, but this constrained version of the theory cannot 
serve to define an acceptable model unless the perturbative solution 
converges.

The technique is to regard higher derivative parts of the 
Euler-Lagrange equation as a perturbation and then use the unperturbed
equation to reduce the order \cite{JLM}. Of course this produces a 
remainder with even more higher derivatives, but this remainder is also higher
order in perturbation theory. By iterating the procedure infinitely, and 
then neglecting the remainder, one obtains a lower order equation. 

The technique can be illustrated for the higher derivative oscillator 
(\ref{HDO}) by regarding the parameter $\epsilon$ as a coupling constant 
so that the Euler-Lagrange equation (\ref{HDE}) takes the form,
\begin{equation}
\ddot{x} + \omega^2 x = -\epsilon \Bigl( \frac{d}{d \omega t}\Bigr)^2 
\ddot{x} \equiv -\epsilon D^2 \ddot{x} \; . \label{step1}
\end{equation}
The first iteration gives,
\begin{equation}
\ddot{x} + \omega^2 x = + \epsilon \ddot{x} + \epsilon^2 D^4 \ddot{x}
= -\epsilon \omega^2 x - \epsilon^2 (1 \!-\! D^2) D^2 \ddot{x} 
\; . \label{step2}
\end{equation}
After another iteration one obtains,
\begin{eqnarray}
\lefteqn{\ddot{x} + \omega^2 x = -\epsilon \Bigl[1 \!+\! \epsilon (1 \!+\! 
\epsilon)^2 (2 \!+\! \epsilon)\Bigr] \omega^2 x } \nonumber \\
& & \hspace{3cm} - \epsilon^4 \Bigl[(2 \!+\! \epsilon)(1\!+\! \epsilon) - 
(2\!+\!\epsilon) D^2 + D^4\Bigr] (1 \!-\! D^2) D^2 \ddot{x} \; . \qquad
\label{step3}
\end{eqnarray}
Continuing in this fashion, {\it and ignoring the remainder}, gives,
\begin{equation}
\ddot{x} + k_+^2 x = 0 \; . \label{pertsol}
\end{equation}
From the full theory, the perturbative development has retained only 
the solution whose frequency is well behaved for $\epsilon \rightarrow 
0$,
\begin{equation}
k_+^2 = \omega^2 \Bigl[1 + \epsilon + 2 \epsilon^2 + 
O(\epsilon^3) \Bigr] \; . \label{lowk}
\end{equation}
It has discarded the solution whose frequency blows up as $\epsilon 
\rightarrow 0$,
\begin{equation}
k_-^2 = \omega^2 \Bigl[\frac1{\epsilon} - 1 - \epsilon - 2 \epsilon^2 
+ O(\epsilon^3)\Bigr] \; . \label{highk}
\end{equation}

The perturbative development (\ref{pertsol}) is what results if one changes
the original theory by imposing the constraints,
\begin{eqnarray}
\ddot{q}(t) = - k_+^2 q(t) \qquad & \Longleftrightarrow & \qquad P_2 = 
\frac{m}2 \Bigl(1 \!-\! \sqrt{1 \!-\! 4\epsilon}\Bigr) X_1 \; , \label{C1} \\
q^{(3)}(t) = - k_+^2 \dot{q}(t) \qquad & \Longleftrightarrow & \qquad X_2 =
\frac1{2 \epsilon m} \Bigl(1 \!-\! \sqrt{1 \!-\! 4\epsilon}\Bigr) P_1 \; . 
\label{C2}
\end{eqnarray}
Under these constraints the Hamiltonian becomes,
\begin{equation}
H_{\rm pert} = \sqrt{1 \!-\! 4\epsilon} \Bigl( \frac{m}2 X_2^2 + 
\frac{m k_+^2}2 X_1^2 \Bigr) \; ,
\end{equation}
which is that of a positive energy harmonic oscillator with mass 
$\sqrt{1 - 4\epsilon} m$ and frequency $k_+$. If the constraints 
(\ref{C1}-\ref{C2}) are imposed at one instant, they remain valid for all 
times as a consequence of the full equation of motion (\ref{HDE}), so
the constrained model is consistent. This is ultimately a consequence of
the fact that, for this model, the perturbatve expansion converges. That
is what ensures that the discarded remainder term really goes to zero 
when the expansion is carried to infinite order. 

For nonlinear Euler-Lagrange equations it is more difficult to reach a
second order form, but one can still do it. As before, the ultimate
consistency of the reduced system depends upon the convergence of the
perturbative expansion. For certain mechanical systems it does converge,
for example, a higher derivative generalization of a particle moving in
a uniform gravitational acceleration $g$ is,
\begin{equation}
L = \frac12 m \dot{x}^2 + m g x + \frac{\epsilon m}{6 g} x \ddot{x}^2
\qquad \Longrightarrow \qquad \ddot{x} = g + \frac{\epsilon}{6 g} 
\ddot{x}^2 + + \frac{\epsilon}{3 g} \frac{d^2}{dt^2} \Bigl( x \ddot{x}
\Bigr) \; . \label{nongrav}
\end{equation}
Reducing to second order transforms the higher derivative corrections
into a distortion of the acceleration,
\begin{equation}
\ddot{x} = \frac{g}{2 \epsilon} \Bigl[1 - \sqrt{1 \!-\! 2 \epsilon}\Bigr] 
= g \Bigl[1 + \frac12 \epsilon + \frac12 \epsilon^2 + O(\epsilon^3)\Bigr] 
\; . \label{reduced}
\end{equation}   
However, there are no known, interacting, $3+1$-dimensional field theories 
for which the perturbative expansion converges. Nor has anyone ever found
a consistent way of imposing constraints which avoids the Ostrogradskian 
instability for an interacting, $(3 + 1)$-dimensional, higher derivative
field theory.     

\section{Conclusions}

Although it was not apparent in 1850, Ostrogradsky's theorem can today
be recognized as the strongest restriction on what sorts of interacting 
local quantum field theories can describe fundamental physics. No
symmetry principle has a broader scope or comparable power. Its 
applications include:
\begin{itemize}
\item{Demonstrating that higher derivative counterterms cannot be a
fundamental solution to the problem of quantum gravity \cite{RPW3};}
\item{Establishing $f(R)$ models as the only metric-based, local and
potentially stable modifications of gravity \cite{RPW4}; and}
\item{Discussing the problems of nonlocal models which can be viewed 
as the limits of an infinite sequence of higher derivatives \cite{EW}.}
\end{itemize}
One should also note the recent generalization by Motohashi and
Suyama of Ostrogradsky's result to Lagrangian-based systems (such as
fermions) whose Euler-Lagrange equations involve an odd number of time 
derivatives \cite{MS}.

\vskip 1cm
\centerline{Acknowledgements}
It is a pleasure to acknowledge correspondence on this subject with K.
Anagnostopoulos, S. Deser, M. Fasiello, T. A. Jacobson, B. P. Kosyakov, 
H. Motohashi and T. Suyama. This work was partially supported by NSF 
grant PHY-1205591 and PHY-1506513, and by the Institute for Fundamental 
Theory at the University of Florida.

\end{document}